# Spatial Structure of the Cooper Pair

Alan M. Kadin*

Princeton Junction, New Jersey 08550, USA

October 11, 2005

**Abstract:**

The Cooper pair is generally analyzed in momentum space, but its real-space structure also follows directly from the BCS theory. It is shown here that this leads to a spherically symmetrical quasi-atomic wavefunction, with an identical "onion-like" layered structure for each of the electrons constituting the Cooper pair, with charge layers ~ 0.1 nm and radius ~ 100 nm for a classic BCS superconductor. This charge modulation induces a corresponding charge modulation in the background ionic lattice, and the attractive interaction between these two opposite charge modulations produces the binding energy of the Cooper pair. This physically-based interaction potential is similar to that in the simple BCS approximation. The implications of this real-space picture for understanding conventional and exotic superconductors are discussed.

*E-mail: amkadin@alumni.princeton.edu

**I. Introduction**

The Cooper pair is the bound state of two electrons that forms the basis for super-conductivity within the Bardeen-Cooper-Schrieffer (BCS) theory [1,2]. It has long been established that the mechanism for pair formation in conventional superconductors is the electron-phonon interaction, although this is rather less clear for exotic materials such as the cuprates. The usual analysis in the BCS theory relies on a momentum-space picture, dealing with interactions between plane-wave electron states. In particular, it is often stated that the Cooper pair consists of a pair of electrons with opposite wavevectors **k** and –**k**, which interact via the exchange of a virtual phonon. But this description is incomplete, and may be somewhat misleading.

The scale of the Cooper pair is well known from BCS theory to be the coherence length $\xi_0 \sim 100$ nm, but a detailed real-space picture of the pair wavefunction is not generally described. Weisskopf [3] pointed out in his tutorial article that,
*"The wave function of a Cooper pair represents a bound S-state, ... analogous to the motion of two nucleons in a deuteron or of the two electrons in the ground state of positronium,"*
and Waldram [4] stated that,
*"The internal part of the pair function corresponds to a spherically symmetrical bound s state of high order; the radial k value of of order $k_F$, and the radius of the state is of order $\xi_0$."*
However, in neither case was this real-space analysis carried much further.

In the present paper, a more complete real-space picture of the Cooper pair is developed, based on the standard equations of the BCS theory. This picture includes not only the electrons themselves, but also the corresponding lattice distortions that move with the electrons and facilitate the pairing. Each pair, consisting of a sphere of radius $\sim \xi_0$ with $\sim$ 1000 concentric charge layers (see Fig. 1), represents a coherent state associated with a particular quantum phase. The density of pairs is such that they are highly overlapping, leading to long-range phase coherence if the quantum phases are all aligned. Finally, this view of the spatial structure of Cooper pairs provides a basis to consider the similarities and differences between conventional BCS superconductors and alternative pairing mechanisms.

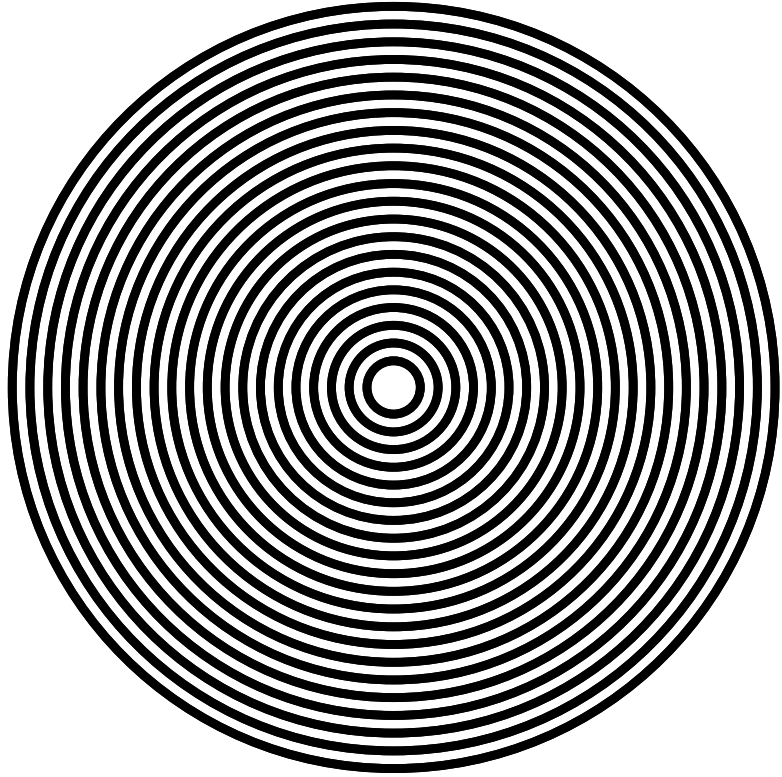

*Fig. 1. Conceptual picture of the spatial structure of a Cooper pair. Spherical standing waves with Fermi wavevector $k_F$ give rise to concentric charge layers with a radius ~ $\xi_0$. Commensurate charge modulation in the underlying lattice serves to bind the two electrons to form the pair.*

**II. Pair Wavefunction**

Consider first the electronic orbitals in a hydrogen atom. The 1S ground state has spherical symmetry and no radial nodes, with a characteristic size given by the Bohr radius. Higher order S states maintain the spherical symmetry, but have multiple radial nodes and wave oscillations, corresponding to spherical standing waves of higher order and greater kinetic energy. The 1000S orbital would have 999 radial nodes. Also consider the bound state of an electron and a positron, also known as positronium. The orbitals of both components are hydrogenic orbitals, each spatially distributed over the same scale (since their masses are the same), symmetric around the common center of mass.

Now consider by analogy a bound state of two electrons of opposite spin, the Cooper pair. One would similarly expect that each of these constituent electrons will form a distributed, spherically symmetrical S-wave orbital. In the ground state, one would also expect that the spatial parts of the two wave functions should be identical. It is also expected that the electron wave components constituting the Cooper pair will have wavevectors near $k_F$, the Fermi wavevector at the top of the Fermi sea. The size of the Cooper pair is inversely proportional to the effective bandwidth of the band of electron components contributing to the pairing. Within the BCS theory, this effective bandwidth also provides self-consistently the binding energy of the Cooper pair.

The BCS ground state at $T=0$ consists of two classes of electrons: those deep inside the Fermi sea, which are essentially the same as those in the normal state, and those near the Fermi surface, which form the superposition of Cooper pairs. These latter electrons, the "frozen crust" at the top of the Fermi sea [3], cannot scatter because they are in a coherent state. The former electrons inside the sea cannot scatter because they are far from the surface. At $T=0$, all electrons of both classes contribute to the lossless supercurrent, but it is the Cooper pair wave function that is of interest here.

The internal structure of this pair wave function $\Psi(r)$ (also called the singlet pair function or the Gor'kov F function) is given by [4]

$$\Psi(r) \propto \sum_k u_k v_k \exp(ikr) \propto \sum_k \frac{\cos(kr)}{\sqrt{\varepsilon^2 + \Delta^2}} = N(0) \int \frac{\cos(kr) d\varepsilon}{\sqrt{\varepsilon^2 + \Delta^2}}, \qquad (1)$$

where the sum is over $k$ states near $k_F$, $\varepsilon = E-E_F$, where $E= \hbar^2 k^2/2m$ and $E_F = \hbar^2 k_F^2/2m$ are the free electron kinetic energy and the Fermi energy, and $\Delta$ is the BCS superconducting energy gap. The functions $u_k$ and $v_k$ are the standard variational parameters of the BCS theory, such that $2u_k v_k = [\varepsilon^2+\Delta^2]^{-1/2}$. The transition to an integral over $\varepsilon$ is possible since only the radial components of $k$ are included, and $N(\varepsilon=0)$ (the density of states for a single electron spin per unit energy at the Fermi surface) can be pulled out of the integral since it is limited to contributions close to $\varepsilon=0$ ($k$ near $k_F$). Note also that the offset energy $\varepsilon$ can be rewritten

$$\varepsilon = \frac{\hbar^2}{2m}(k^2 - k_F^2) \approx \frac{\hbar^2 k_F}{m}(k - k_F) = \pi \xi_0 \Delta (k - k_F), \qquad (2)$$

where we have defined the BCS coherence length $\xi_0 = \hbar v_F/\pi\Delta$ in the usual way, and the approximation is valid for $k$ close to $k_F$. Now one can define normalized parameters $\varepsilon' = \varepsilon/\Delta$ and $r' = r/\pi\xi_0$, and Eq. (1) becomes

$$\Psi(r) \propto \int \frac{\cos(k_F r + \varepsilon' r')d\varepsilon'}{\sqrt{\varepsilon'^2 + 1}} \approx \cos(k_F r)\int \frac{\cos(\varepsilon' r')d\varepsilon'}{\sqrt{\varepsilon'^2 + 1}} = \cos(k_F r)K_0(r/\pi\xi_0), \quad (3)$$

Here $K_0$ is the zeroth-order modified Bessel function, with an asymptotic form that is similar to an exponential: $K_0(x) \sim (\pi/2x)^{1/2}\exp(-x)$ for $x \gg 0$ [5]. The $K_0$ function has a weak divergence for $r=0$, which must be cut off by choosing a cutoff energy scale. In the BCS theory, this cutoff is given by an energy comparable to the Debye energy $\hbar\omega_D$, which is $\gg \Delta$.

As shown in Fig. 2a, the Fourier component peaks at $k_F$ with a width $\sim 2\Delta$, as expected. So this wavefunction corresponds to spherical standing waves, with a spatial modulation close to $k_F$. Given that typically $1/k_F \sim 0.1$ nm and $\xi_0 \sim 100$ nm, this corresponds to a rapidly oscillation around $k_F$, modulating a slowly varying envelope with a characteristic scale of $\xi_0$. This is illustrated in Fig. 2b, where the factor $k_F\xi_0 = 20/\pi$ for clarity. (Calculations are carried out using Matlab®, in which $K_0$ is a built-in function.)

The charge density in this electron wavefunction goes as the square of the wavefunction:
$$\rho(r) \propto K_0^2(r/\pi\xi_0) \cos^2(k_F r) \propto [1 + \cos(2k_F r)] \exp(-2r/\pi\xi_0)/r, \quad (4)$$
where the latter form is the asymptotic expression for large $r$. Both electrons of the pair would be expected to have the same (spatial) wavefunction, with the same nodes and antinodes, and hence the same quasi-static charge distribution. This corresponds to a spherical layered charge distribution of the Cooper pair, with periodic layers spaced by $2\pi/2k_F \sim 2$ Å. If the Cooper pair moves, the charge distribution will move with it.

The average charge of the slowly varying envelope is easily screened and probably not important. But the charge modulation at $2k_F$, which is similar to that of a high-order atomic S-state, induces a corresponding modulation in the underlying lattice (described further in the next section), and the attraction between the two commensurate modulations provides the physical basis for the BCS bound state.

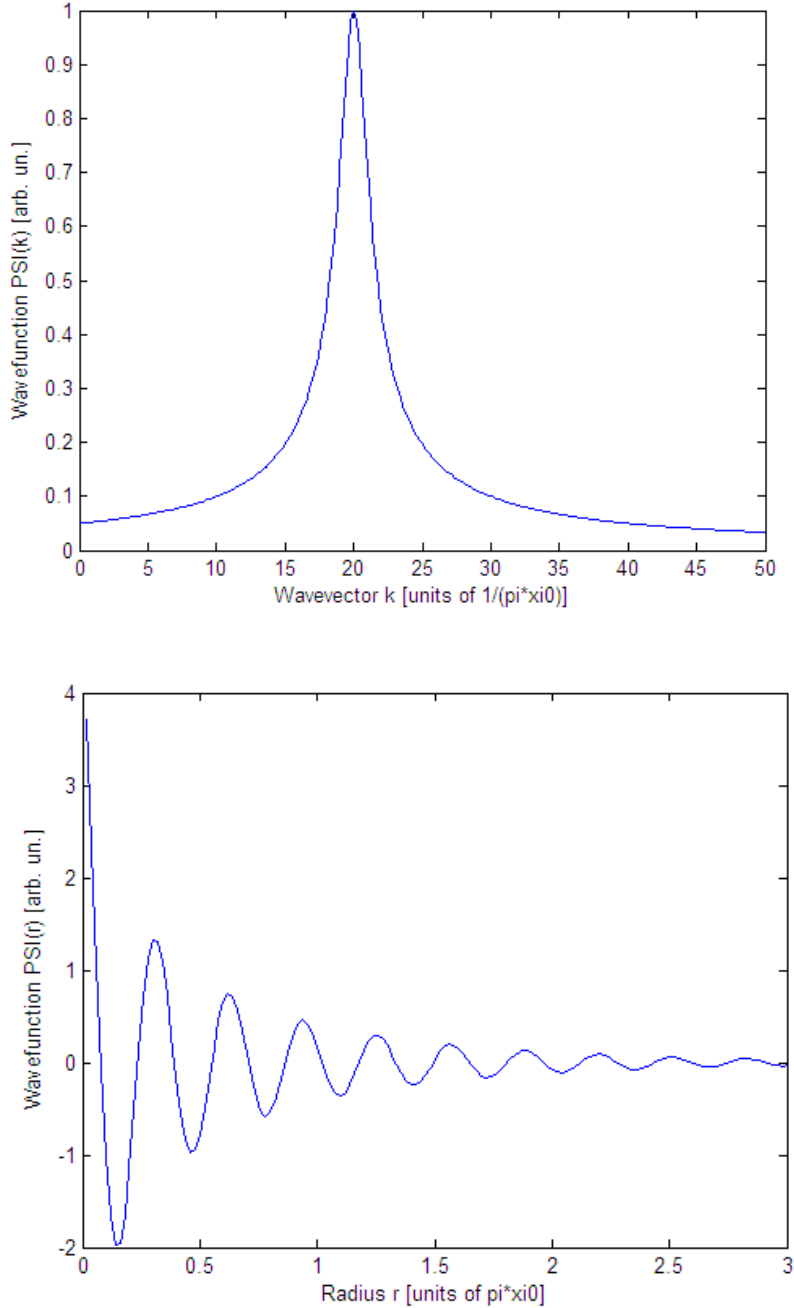

*Fig. 2. Cooper pair wavefunction in real and reciprocal space. For clarity, $k_F=20/\pi\xi_0$ ($E_F=10\Delta$) in these figures. (a) K-space distribution showing peak at $k_F$ and peak width $2/\pi\xi_0$ (corresponding to $2\Delta$ in energy). (b) Radial dependence of spherical standing wave, showing modified Bessel function $K_0$ with quasi-exponential decay over distance ~ $\pi\xi_0$. Calculations are carried out using Matlab®.*

The effective radius of a Cooper pair can be estimated by evaluating the mean radius of the envelope of the charge distribution:

$$\langle r \rangle = \frac{\int 4\pi r^2 \rho(r) r \, dr}{\int 4\pi r^2 \rho(r) \, dr} \approx \pi \xi_0, \tag{5}$$

where we have used the asymptotic form for the calculation. This is somewhat larger than $\xi_0$, which is typically taken as the scale of the Cooper pair. It is notable that the Cooper pairs are heavily overlapping, as one can see by estimating the number of Cooper pairs in a volume of radius $\pi\xi_0$. Taking the effective energy bandwidth of electrons contributing to the Cooper pair as $2\Delta$, the number $n$ of pairs in such a volume, assuming that all electrons within this band are paired (as is the case far below the superconducting critical temperature) is:

$$n = N(0)\, 2\Delta\, (4\pi/3)(\pi\xi_0)^3 \approx 1.7\, (E_F/\Delta)^2, \tag{6}$$

using formulas for the free-electron Fermi gas. For conventional low-temperature superconductors, $E_F \sim 1$ eV while $\Delta \sim 1$ meV, so that there are $\sim 10^6$ Cooper pairs overlapping within an effective volume of a given Cooper pair.

In the absence of electromagnetic fields, each Cooper pair consists of two identical electron wave functions (with opposite spins), each at rest with zero net momentum. Although the sign of the electron wavefunction alternates on the scale of $1/k_F$, there is an overall quantum phase factor $\phi$ for the entire Cooper pair which is essentially uniform. If the magnetic vector potential $\mathbf{A} \neq 0$, the pair is in motion with velocity $\mathbf{v_s} = 2e\mathbf{A}/2m = e\mathbf{A}/m$. The assembly of overlapping Cooper pairs all move together, each with the same quantum energy, and hence the same time-evolution of the quantum phase $\phi$. Clearly, this superconducting condensate exhibits long-range coherence only if all of the overlapping Cooper pairs are in-phase.

### III. Lattice Modulation

Within the BCS theory, the lattice potential energy for the electrons is described in reciprocal space as $V_{kk'} = V(k-k') = V(q)$, where $k$ and $k'$ are the wavevectors of two electron states that are coupled by the potential. Since the coupled states are on opposite sides of the Fermi surface, one expects that values of $q$ near $2k_F$ are most relevant to the pairing. For simplicity of calculations, the BCS approximation takes $V_{kk'}$ as a constant $-V$ for states within a cutoff energy $\hbar\omega_c$ of the Fermi surface, and zero for all other states [2]. On physical grounds, $\hbar\omega_c$ is taken to be of order of the Debye energy $\hbar\omega_D$ of the phonons, at least for phonon-mediated superconductivity. The Debye energy is intermediate in magnitude between $E_F$ and the superconducting energy gap $\Delta$.

Within the present real-space picture of the Cooper pair, one may view $V(q)$ for $q=2k_F$ as coupling together the outgoing and incoming traveling-wave components of the radial standing waves for both of the electrons. It is reasonable to identify $V(q)$ with the electric potential due to the charge self-consistently induced in the lattice by the electron wavefunction, which also is predominantly around $2k_F$. Given the spherical symmetry and the narrow bandwidth, this is essentially a one-dimensional problem, with Fourier transform $V(r)$ given by

$$V(\mathrm{r}) \propto -V \cos(2k_F r)\, \sin(2k_c r)/(2k_c r), \qquad (7)$$

where $k_c = (\hbar\omega_c/\Delta)/\pi\xi_0$ corresponds to the energy cutoff at $\hbar\omega_c$. As shown in Fig. 3, this corresponds to a modulated potential which generally decreases slowly ($\sim 1/r$) in amplitude, but this envelope also oscillates as it dies out. These envelope oscillations (and nodes) appear to be a mathematical artifact of the sharp cutoff at $\hbar\omega_c$, rather than being physically significant.

This is similar, but not identical to a more physically-based potential that follows from the electronic charge distribution of Eq. (4). One might initially expect that this modulated charge density would be screened by other electrons, but in fact all of the conduction electrons that can contribute at this wave vector are already included in the superconducting condensate. Thus the only remaining effect is due to the dielectric response of the bound ions. The modulated spherical shells of charge given by Eq. (4) would be expected to induce a proportional modulated charge $\rho_{in}$ in the lattice, of opposite phase, reducing the total charge density. The contribution to $V(r)$ due to a spherical shell of charge $Q$ is $Q(r)/4\pi\varepsilon_0 r$, and since $Q(r)$ oscillates quickly, one would expect that only the shell nearest $r$ would contribute significantly. Then one has

$$V(r) \propto Q(r)/r \propto r\rho_{in}(r) \propto \cos(2k_F r)\exp(-2r/\pi\xi_0) \qquad (8)$$

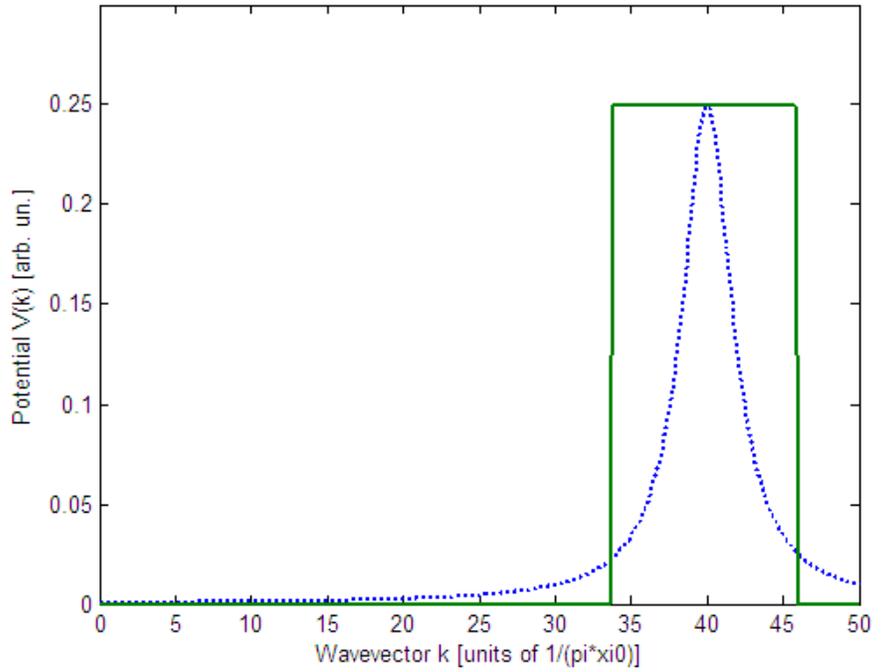

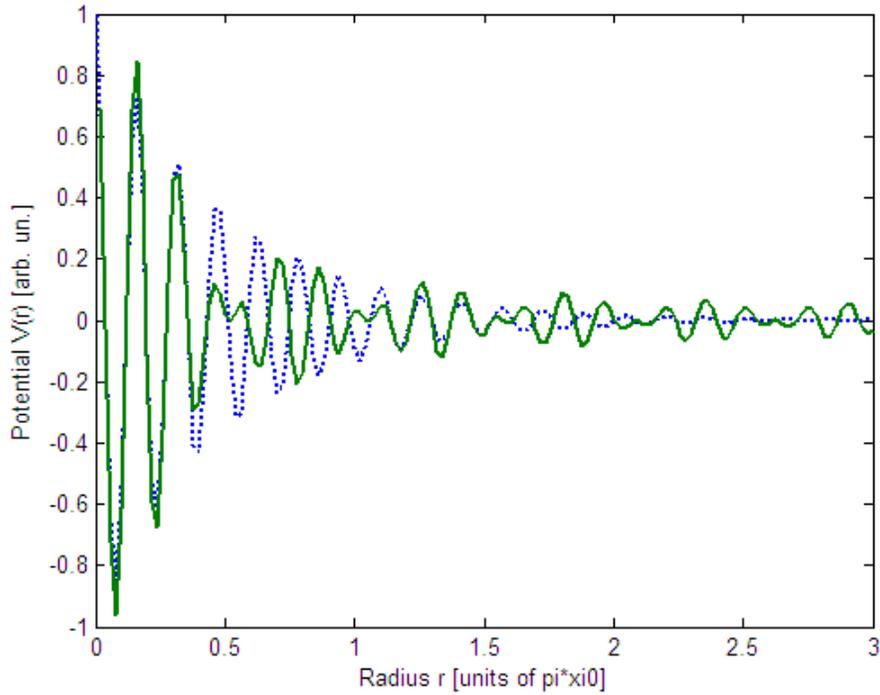

*Fig. 3. Potential energy of Cooper pair associated with lattice. Both constant BCS approximation (solid lines) and charge-induced lattice modulation (dotted lines) are shown. For clarity, $k_F=20/\pi\xi_0$ ($E_F=10\Delta$) in these figures, and the cutoff energy $\hbar\omega_c=3\Delta$. (a) K-space distribution showing peak at $2k_F$ and peak width corresponding either to $2\Delta$ or $2\hbar\omega_c$. (b) Radial dependence of potentials, showing modulated wave at $2k_F$ with either exponential rolloff or sin(x)/x oscillating rolloff.*

One can also view the Fourier transform of this:

$$V(k) \propto [(k-2k_F)^2 + (2/\pi\xi_0)^2)]^{-1} \qquad (9)$$

These dependences are also shown in Fig. 3, for comparison with that of the BCS approximation. Note that the potential in Eq. (8) exhibits the same fine modulation at $2k_F$ as that of Eq. (7), but the envelope decreases more smoothly to zero at $\sim\pi\xi_0$, without further oscillations at larger distances.

Note that either form of the lattice distortion is static for a Cooper pair that is fixed in space. This seems quite different from the conventional picture of a dynamically oscillating phonon, but may properly represent a "virtual phonon" that is bound to the Cooper pair. Also, since the lattice distortion moves with the Cooper pair, a moving Cooper pair does indeed correspond to dynamic lattice vibrations. The maximum velocity of a Cooper pair at T=0 [6] is given by $v_c = \hbar/(\pi m \xi_0) = \Delta/\hbar k_F$, so that the maximum frequency of the lattice oscillations is

$$f_{max} = v_c k_F/\pi = 2\Delta/h. \qquad (10)$$

Since $2\Delta$ is typically much less than characteristic phonon energies of order the Debye energy, this is quite consistent with a phonon-mediated picture of superconductivity.

### III. Discussion and Conclusions

The analysis developed here provides an alternative real-space picture for a Cooper pair within the standard BCS model of an s-wave superconductor. However, it may be extended to cover superconductors with other symmetries and other pairing mechanisms. For example, there has been considerable discussion of d-wave pairing symmetry as applied to the cuprates [7], as well as p-wave symmetry in ruthenates [8]. It is straightforward to modify the spherically symmetric quasi-atomic orbital of Fig. 1, to include one or more angular nodes, as shown in Fig. 4, by multiplying the wavefunction by $\cos\phi$ or $\cos(2\phi)$. This corresponds to rotational standing waves as well as radial standing waves, and leads to separate lobes of the wavefunction that are 180° out of phase with adjacent lobes, as for the usual case with atomic orbitals. Hybrid orbitals are standard in atomic bonding (such as $sp^2$ or $sp^3$, for example), and a pair structure that is likewise hybridized may also be a possibility.

While such pictures do not specify the nature of the pairing mechanism, they imply that these alternative symmetries might be favored in certain anisotropic crystal structures, where lattice modulations in certain spatial directions might be suppressed. This may be relevant to the cuprates, for example, where the directions of the d-wave lobes are believed to correspond to the directions of the Cu-O bonds within the a-b plane of the crystal structure.

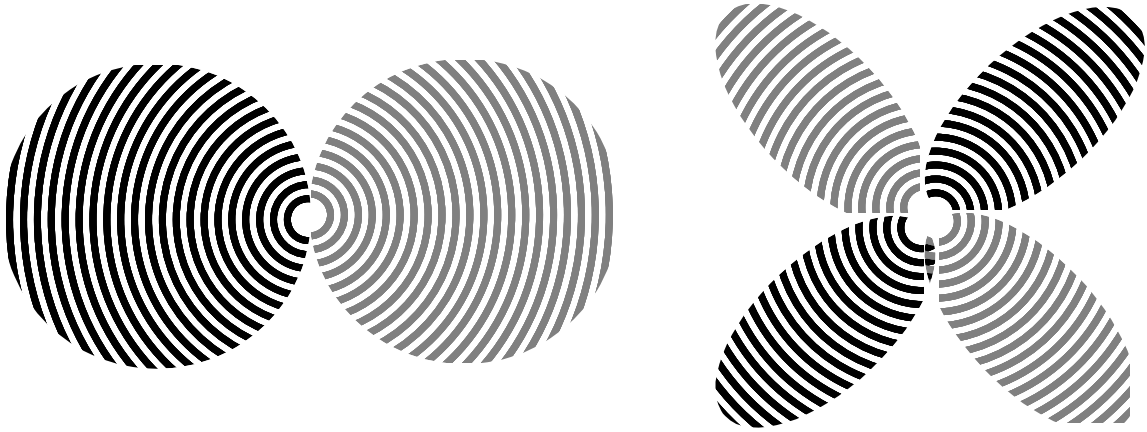

*Fig. 4. Conceptual picture of Cooper pair with alternative p-wave and d-wave symmetries (left and right). In addition to the layered structure with radial nodes, one would have angular nodes as well, corresponding to p and $d_{x^2-y^2}$ atomic orbitals. When one crosses an angular node, the quantum phase changes by 180°, as indicated by the change in shading of the various lobes.*

In describing the physical picture of Cooper pair bonding via induced charge modulation, we asserted earlier that only the lattice of ions is available to respond to the charge modulation in the pair wavefunction. However, one can also envision a more general situation whereby bound orbitals in other electron bands or sublattices may also be available to respond at $2k_F$. This might correspond to certain models of excitonic superconductivity or other exotic forms.

Furthermore, we assumed earlier that both electrons of opposite spin have overlapping identical wavefunctions. This has the effect of reinforcing the charge modulations, while canceling out any spin density modulations. One could alternatively cancel the charge modulations (by having the radial nodes of one electron align with antinodes of the other), which would have the effect of reinforcing the spin density modulations. If one has a magnetically polarizable lattice (such as is believed to be the case for the cuprates and some other exotic superconductors), then such a spin density modulation could induce a corresponding modulation in the spin of the lattice, and the interaction between the two spin modulations might provide the binding energy of the Cooper pair. This may be similar to certain models that have been proposed to account for high-temperature superconductivity in the cuprates. Alternatively, one could have a "spin-triplet" state whereby the two electrons have parallel spins, as is believed to be the case for ruthenate superconductors [8]. With identical aligned wavefunctions, one would then have a modulation of both charge density and spin density.

Another possible extension of this picture would permit coupling to dynamic lattice distortions, as opposed to the quasi-static distortions that follow from the simple model described. For example, spherical standing waves in the lattice at a phonon frequency $\omega(2k_F)$ would produce charge oscillations at the same frequency and wavevector. Since electron wavefunctions can respond much faster than the lattice, a Cooper pair could couple electrostatically to such a "bound phonon" simply by a slight breathing oscillation of the entire spherical wavefunction at the resonant frequency.

Finally, the hallmark of superconductivity is the long-range order associated with the quantum phase $\phi$. This is generally characterized using the Ginzburg-Landau phenomenological theory, in which an order parameter $|\Psi| \exp(i\phi)$ represents an effective wavefunction for the overlapping Cooper pairs. This changes slowly on the scale of an empirical coherence length $\xi(T)$ that varies with temperature $T$. It is reasonable to associate $\xi(T)$ with the BCS coherence length $\xi_0$, which is why they are indicated by the same Greek letter. In a "clean superconductor", where the electron mean-free-path $\ell > \xi_0$, one indeed has $\xi(T) \sim \xi_0$. However, in a disordered or "dirty superconductor" with $\ell < \xi_0$, the situation is a bit more complicated [9]. One has instead $\xi(T) \sim (\xi_0\ell)^{1/2}$, which seems inconsistent with a Cooper pair on the scale of $\xi_0$. In fact, the pair wavefunction must be modified in the dirty limit, since the basis electron wavefunctions on this scale are no longer plane waves. Instead of ballistic motion in a straight line, the electron undergoes a random walk between elastic scattering centers. So an energy bandwidth on the scale of $\Delta$ will lead to a smaller Cooper pair on the scale of $(\xi_0\ell)^{1/2}$, with an internal microstructure that is more complicated than the simple layered structure of Fig. 1. However, one would still expect a spatial charge modulation and corresponding induced lattice charge, leading to a coherent wavefunction with a pairing energy that is essentially independent of $\ell$.

In conclusion, a simple real-space physical picture for the Cooper pair is developed, which is consistent with the standard equations of BCS theory. This Cooper pair consists of a spherical quasi-atomic wavefunction of radius $\sim \pi\xi_0$ for both electrons, with radial nodes separated by $\pi/k_F$. A commensurate quasi-static charge modulation in the underlying lattice provides the interaction potential, similar to that in the simple BCS approximation, that leads to a bound state. This picture can be generalized for alternative pairing symmetries (p and d-wave) as well as alternative pairing mechanisms (excitonic or spin-based) and extended to disordered superconductors. While it does not derive essentially new physics, this real-space approach may provide a more effective heuristic picture for developing and understanding new phenomena in superconductivity.

**References:**
[1] L.N. Cooper, "Bound Electron Pairs in a Degenerate Fermi Gas," *Phys. Rev.* **104**, 1189 (1956).
[2] J. Bardeen, L.N. Cooper, and J.R. Schrieffer, "Theory of Superconductivity," *Phys. Rev.* **108**, 1175 (1957).
[3] V.F. Weisskopf, "The Formation of Superconducting Pairs and the Nature of Superconducting Currents," *Contemporary Physics* 22, 375 (1981).
[4] J.R. Waldram, *Superconductivity of Metals and Cuprates*, Section 9.3 (Institute of Physics, Bristol, 1996).

[5] I.S.Gradshteyn and I.M. Ryzhik, *Tables of Integrals, Series and Products* (Academic Press, New York, 1980).

[6] M. Tinkham, *Introduction to Superconductivity*, 1st ed., p. 119 (McGraw-Hill, New York, 1975).

[7] C.C Tsuei and J.R. Kirtley, "Pairing Symmetry in Cuprate Superconductors," *Rev. Mod. Phys*. **72**, 969 (2000).

[8] A.P. MacKenzie and Y. Maeno, "The Superconductivity of $Sr_2RuO_4$ and the Physics of Spin Triplet Pairing," *Rev. Mod. Phys*. **75**, 657 (2003).

[9] J.R. Waldram, *Superconductivity of Metals and Cuprates*, p. 191 (Institute of Physics, Bristol, 1996).